\documentclass[prd,twocolumn,showpacs,nofootinbib]{revtex4}

\usepackage{amsmath}
\usepackage{graphicx}
\usepackage{dcolumn}
\usepackage{bm}
\usepackage{amssymb}
\usepackage{latexsym}
\usepackage{subfigure}

\usepackage{natbib}
\usepackage{bm}

\usepackage{graphicx}
\usepackage{amsmath}
\usepackage{amssymb}
\usepackage{hyperref}
\usepackage{verbatim}

\newcommand{\nilc}{{\tt NILC\,}}
\newcommand{\sevem}{{\tt SEVEM\,}}
\newcommand{\smica}{{\tt SMICA\,}}
\newcommand{\commander}{{\tt COMMANDER\,}}

\def \der{{\rm d}}

\def\kms{\,{\rm km}\,{\rm s}^{-1}}

\def \r1r2{{|{\bf r}_{1}-{\bf r}_{2}|}}
\def \fgas{{{f}_{\rm gas}}}

\begin{document}

\title{Constraining the ionized gas evolution with CMB-spectroscopic survey cross-correlation}

\author{Yin-Zhe Ma$^{1,2}$}

\affiliation{$^{1}$School of Chemistry and Physics, University of KwaZulu-Natal, Westville Campus, Private Bag X54001, Durban, 4000, South Africa}
\affiliation{$^{2}$NAOC--UKZN Computational Astrophysics Centre (NUCAC), University of KwaZulu-Natal, Durban, 4000, South Africa}

% didn't use \today because this annoyingly changes the date every time

\begin{abstract}
We forecast the prospective constraints on the ionized gas model $f_{\rm gas}(z)$ at different evolutionary epochs via the tomographic cross-correlation between kinetic Sunyaev-Zeldovich (kSZ) effect and the reconstructed momentum field at different redshifts. The experiments we consider are the {\it Planck} and CMB Stage-4 survey for CMB and the SDSS-III for the galaxy spectroscopic survey. We calculate the tomographic cross-correlation power spectrum, and use the Fisher matrix to forecast the detectability of different $f_{\rm gas}(z)$ models. We find that for constant $f_{\rm gas}$ model, {\it Planck} can constrain the error of $f_{\rm gas}$ ($\sigma_{f_{\rm gas}}$) at each redshift bin to $\sim 0.2$, whereas four cases of CMB-S4 can achieve $\sigma_{f_{\rm gas}} \sim 10^{-3}$. For $f_{\rm gas}(z)=f_{\rm gas,0}/(1+z)$ model the error budget will be slightly broadened. We also investigate the model $f_{\rm gas}(z)=f_{\rm gas,0}/(1+z)^{\alpha}$. {\it Planck} is unable to constrain the index of redshift evolution, but the CMB-S4 experiments can constrain the index $\alpha$ to the level of $\sigma_{\alpha} \sim 0.01$--$0.1$. The tomographic cross-correlation method will provide an accurate measurement of the ionized gas evolution at different epochs of the Universe. 
  
\end{abstract}

\maketitle

\section{Introduction}
\label{sec:intro}
The measurement of the cosmic microwave background radiation (CMB) temperature anisotropy from {\it Planck} satellite and other large-scale structure measurement (e.g. Type-Ia supernovae, Baryon Acoustic Oscillation from SDSS) have found that the baryonic matter accounts for $4.8\%$ of the total Universe' budget~\cite{Planck15-para}. However, by counting the amount of baryons in form of stars, interstellar medium and intracluster medium, there are more than $50$ per cent of the baryons is still missing~\cite{Fukugita04}. Searching for the missing baryons is a crucial step towards fully understanding of galaxy formation, and the interplay between dark matter, baryons and gravity. Hydrodynamic simulation shows that the majority of baryons are diffuse among the intergalactic medium(IGM) with temperature in between $10^{5}\,{\rm K}<T<10^{7}\,{\rm K}$, namely warm-hot intergalactic medium (WHIM)~\cite{Dave01,Cen06}. Because of its temperature range, the WHIM emitted radiation in the UV and soft X-ray bands is too weak to be detected~\cite{Bregman07}. A complementary study is to cross-correlate the thermal Sunyaev-Zeldovich (tSZ)~\cite{Sunyaev72,Sunyaev80} effect with weak gravitational lensing~\cite{Waerbeke14,Ma15,Hojjati15,Hojjati16}. But since tSZ effect is sensitive to the pressure of the gas ($P_{\rm e}$), one needs to separate the temperature of the WHIM in order to infer its density distribution. In fact, the recent studies~\cite{Hojjati15,Hojjati16} showed that such studies are more sensitive to the AGN feedback mechanism in the galaxy clusters than inferring baryon density.

There have been a lot of effort of using kinetic Sunyaev-Zeldovich effect (hereafter kSZ, ~\cite{Sunyaev72,Sunyaev80}) to infer the baryon density around the galaxies and dark matter halos. The kSZ effect describes the temperature anisotropy of the CMB due to the scattering off a cloud of electrons with non-zero peculiar velocities with respect to the CMB rest frame, i.e. 
\begin{eqnarray}
\frac{\Delta T}{T}(\mathbf{\hat{n}})=-\frac{\sigma_{\rm T}}{c} \int n_{\rm e} \left(\mathbf{v}\cdot \mathbf{\hat{n}} \right) \der l ,\label{eq:kSZ1}
\end{eqnarray}
where $\sigma_{\rm T}$ is the Thomson cross-section, $n_{\rm e}$ is the electron density, ($\mathbf{v}\cdot \mathbf{\hat{n}}$) is the velocity along the line-of-sight, and $\der l$ is the integral on the radial direction. Ref.~\cite{Hand12} applied the pairwise momentum estimator (the estimator quantifying the difference in temperature between a pair of galaxies) to the kSZ map observed by Atacama Cosmology Telescope (ACT) and obtained the first detection in 2012. Furthermore, Ref.~\cite{Planck16-unbound} solidified the detection by applying the pairwise momentum estimator to 
{\it Wilkinson Microwave Anisotropy Probe} ({\it WMAP}) 9-year W-band data, {\it Planck} foreground cleaned \sevem, \smica, \nilc, and \commander maps, and the measurements are at a $3.3\sigma$ and $1.8$--$2.5\sigma$ confidence level (CL) for {\it WMAP} and {\it Planck} respectively. In addition, Ref.~\cite{Planck16-unbound} reconstructed the linear velocity field with Central Galaxy Catalog (CGC) selected from Sloan Digital Sky Survey's Data Release 7 (SDSS-DR7), and cross-correlated the {\it Planck}'s kSZ field with velocity field ($\langle \Delta T (\mathbf{v}\cdot \mathbf{\hat{n}}) \rangle$). It found the detection at $3.0$--$3.2\sigma$ CL for the foreground cleaned {\it Planck} maps (namely, \sevem, \smica, \nilc, and \commander \,maps), and $3.8\sigma$ CL for the {\it Planck} 217 GHz raw map. A following-up paper~\cite{Carlos} showed that the measured value of optical depth ($\tau=(1.39 \pm 0.46) \times 10^{-4}$) indicates that essentially all baryons are tracing underlying dark matter distribution. More recently, the squared kSZ fields from {\it WMAP} and {\it Planck} were cross-correlated with the projected galaxy overdensity from {\it Wide-field Infrared Survey Explorer} ({\it WISE}) which leaded to $3.8\sigma$ CL detection. With advanced ACTPol and hypothetical Stage-IV CMB experiment the signal-to-noise ratio of the kSZ squared field and projected density field can reach $120$ and $150$ respectively~\cite{Ferraro16}. By cross-correlating the velocity field from CMASS samples with the kSZ map produced from ACT observation, Ref.~\cite{Schaan16} detected the aggregated signal of kSZ at $\sim 3.3\sigma$ CL. In addition, Ref.~\cite{Bernardis16} applied the pairwise momentum estimator to the ACT data and $50,000$ bright galaxies from BOSS survey, and obtained $3.6 \sigma$--$4.1\sigma$ CL detection. By using the pairwise momentum estimator to the South Pole Telescope (SPT) data and Dark Energy Survey (DES) data, Ref.~\cite{Soergel} obtained the averaged central optical depth of galaxy cluster at $4.2\sigma$ CL.

In spirit of constraining the baryon content, Ref.~\cite{Planck16-unbound} presented the method of cross-correlating the kSZ map with the reconstructed velocity field. In fact, one can do the tomographic kSZ measurement at different redshift bins for the very deep spectroscopic survey such as SDSS-III, SDSS-IV and BOSS surveys. Ref.~\cite{Shao11} discussed this method by considering future cross-correlation between {\it Planck} and BigBOSS. Ref.~\cite{Shao16} discussed the prospects of using this cross-correlation technique to constrain the electron density profile of galaxies. In this work, we will forecast the prospective scope of measurement of cosmic ionized gas fraction by doing the tomographic cross-correlation of kSZ with optical survey. We will consider the {\it Planck} survey and the four cases of CMB Stage-4 surveys in the future. For optical survey, we consider the SDSS-III survey as laid out in Table~\ref{tab-sample}. The aim of this paper is to provide a forecast of the precision for future experiments to constrain the total gas fraction of the Universe, therefore provides the inference of the missing baryons.

This paper is organized as follows. In Sect.~\ref{sec:method} we calculate the kSZ tomography, the reconstructed momentum field template, and its cross-correlation. We also discuss the different models of $f_{\rm gas}$ evolution, and the Fisher matrix method to forecast the experimental error. In Sect.~\ref{sec:obs}, we discuss the prospective observational data that can be used for the cross-correlation study. In Sect.~\ref{sec:results}, we present the results of the forecast and discuss its implication. The conclusion remark will be in the last Section.

Throughout the paper, except for the $f_{\rm gas}$ models we vary we will use the {\it Planck} 2015 best-fitting cosmological parameters for the spatially flat $\Lambda$CDM cosmology model~\cite{Planck15-para}, i.e. $\Omega_{\rm m}=0.309$; $\Omega_{\Lambda} = 0.691$; $n_{\rm s} = 0.961$; $\sigma_{8} = 0.809$; and $h = 0.68$, where the Hubble constant is $H_{0} = 100 h \kms\,{\rm Mpc}^{-1}$.

\section{Methodology}
\label{sec:method}
\subsection{kSZ tomography}
\label{sec:kSZ-tomo}
We want to do the tomographic cross-correlation between {\it Planck} kSZ and SDSS-III reconstructed momentum field. So we define the kSZ effect at each redshift bin as
\begin{eqnarray}
\Delta_{i}=\left(\frac{\Delta T}{T} \right)_{i},
\end{eqnarray}
which is at comoving distance bin $[\chi_{i}-\Delta \chi_{i}/2, \chi_{i} + \Delta \chi_{i}/2]$. The total kSZ effect is
\begin{eqnarray}
\left(\frac{\Delta T}{T} \right)=\sum_{i}\Delta_{i}.
\end{eqnarray}

From Eq.~(\ref{eq:kSZ1}), the kSZ effect at each redshift bin  is
\begin{eqnarray}
\Delta_{i}=-\frac{\sigma_{\rm T}}{c} \int^{{\chi_{i}+\Delta \chi_{i}/2}}_{\chi_{i}-\Delta \chi_{i}/2}\frac{\der \chi}{1+z}{\rm e}^{-\tau(z)}n_{\rm e}(z)\mathbf{v}\cdot \mathbf{\hat{n}}.
\end{eqnarray}
\begin{eqnarray}
n_{\rm e}(z)=\overline{n}_{\rm e,i}(z)(1+\delta),
\end{eqnarray}
where $\overline{n}_{\rm e,i}$ is the mean ionized electron density at redshift $z$, which is~\cite{Ma14}
\begin{eqnarray}
\overline{n}_{\rm e,i}(z)=\frac{\chi_{\rm e}\rho_{\rm g}(z)}{\mu_{\rm e}m_{\rm p}}=\frac{\chi_{\rm e}\rho_{\rm cr,0}\Omega_{\rm b}}{\mu_{\rm e}m_{\rm p}}f_{\rm gas}(z)(1+z)^{3},
\end{eqnarray}
where $\mu_{\rm e}=1.14$ is the mean weight per electron, $\rho_{\rm cr,0}$ is the critical density at present time. $\chi_{\rm e}$ is the mean electron fraction, which is 
\begin{eqnarray}
\chi_{\rm e}=\frac{1-Y_{\rm p}\left(1-N_{\rm H_{e}}/4 \right)}{1-Y_{\rm p}/2},
\end{eqnarray}
where $Y_{\rm p}=0.24$ is the primordial helium abundance. The $N_{\rm H_{e}}=0$, $1$, $2$ correspond to the none, singly, and doubly ionized helium, for which $\chi_{\rm e}=0.86$, $0.93$ and $1$ correspondingly. Here we assume all helium are ionized so $\chi_{\rm e}=1$. $f_{\rm gas}(z)$ is the fraction of baryons in form of gas, which is the function we want to fit for the tomographic kSZ measurement. 

We further define $\mathbf{p} \equiv (1+\delta)\mathbf{v}$ as the momentum field. Therefore, 
\begin{eqnarray}
\Delta_{i} &=& - \left(\frac{\sigma_{\rm T}\chi_{\rm e}\rho_{\rm cr,0}\Omega_{\rm b}}{\mu_{\rm e}m_{\rm p}c}\right) \nonumber \\
& \times & \int^{{\chi_{i}+ \Delta \chi_{i}/2}}_{\chi_{i}-\Delta \chi_{i}/2}\der \chi \fgas(z) (1+z)^{2}{\rm e}^{-\tau(z)} (\mathbf{p}\cdot \mathbf{\hat{n}}), \label{eq:deltaTT}
\end{eqnarray}
where optical depth to redshift $z$ is
\begin{eqnarray}
\tau(z)=\sigma_{\rm T}\int^{z}_{0}\frac{\overline{n}_{\rm e}(z)}{1+z}\der \chi.
\end{eqnarray}

We can also write the integral as
\begin{eqnarray}
\Delta_{i} &=& - \left(\frac{\sigma_{\rm T}\chi_{\rm e}\rho_{\rm cr,0}\Omega_{\rm b}}{\mu_{\rm e}m_{\rm p}c}\right) \nonumber \\
& \times & \int^{{\chi_{i}+ \Delta \chi_{i}/2}}_{\chi_{i}-\Delta \chi_{i}/2}\der \chi f_{\rm gas}(z) W_{\rm kSZ}(z) (\mathbf{p}\cdot \mathbf{\hat{n}}), \label{eq:deltaTT2}
\end{eqnarray}
where we have defined the kSZ kernel as $W_{\rm kSZ}(z)=(1+z)^{2}{\rm e}^{-\tau(z)}$.

\subsection{Reconstructed momentum field}
\label{sec:momentum}
We want to cross-correlate the kSZ template with the reconstructed momentum field from spectroscopic surveys. In optical survey we observe a density field at a given redshift range, we can always calculate the Fourier mode of velocity field, and then reconstruct the 3D momentum field in the observation, i.e.
\begin{eqnarray}
\mathbf{p}^{\rm rec}=(1+\delta^{\rm rec}(\mathbf{x}))\mathbf{v}^{\rm rec},
\end{eqnarray}
from which one can integrate and calculate the projected momentum effect, i.e.
\begin{eqnarray}
\tilde{\Delta}_{i}(\mathbf{\hat{n}})= \left(\frac{\sigma_{\rm T}\chi_{\rm e}\rho_{\rm cr,0}\Omega_{\rm b}}{\mu_{\rm e}m_{\rm p}c}\right)  \int^{{\chi_{i}+ \Delta \chi_{i}/2}}_{\chi_{i}-\Delta \chi_{i}/2}\der \chi W_{\rm kSZ}(z) (\mathbf{p}\cdot \mathbf{\hat{n}}), \nonumber  \\
\label{eq:delta-i-n}
\end{eqnarray}
note that in the above integral we do not have the $f_{\rm gas}(z)$ component. 

Then we to the Fourier transformation, and calculate the power spectrum of kSZ--reconstructed momentum field cross-correlation. In the following we denote this correlation function at the $i$th redshift bin as $C^{\rm Tp}_{\ell}(\chi_{i})$, which is found to be (all necessary steps of calculation are presented in Appendix~\ref{sec:power-cell}, see also~\cite{Kaiser92})
\begin{eqnarray}
C^{\rm Tp}_{\ell}(\chi_{i}) &=& -\frac{1}{2} f_{\rm gas,i} \left(\frac{\sigma_{\rm T}\chi_{\rm e}\rho_{\rm cr}\Omega_{\rm b}}{\mu_{\rm e}m_{\rm p}c} \right)^{2} \left(\frac{W_{\rm kSZ}(\chi_i)}{\chi_{i}} \right)^{2}  \nonumber \\
& \times &
 \int^{\chi_{i}+\Delta \chi_{i}/2}_{\chi_{i}-\Delta \chi_{i}/2} \der \chi P_{\rm B}\left(\frac{\ell +1/2}{\chi} \right) \label{eq:Cell-Tp}, 
\end{eqnarray}
where
\begin{eqnarray}
P_{\rm B}\left(k,z \right) &=& (\dot{a}f)^{2} \int \frac{\der^{3}\mathbf{k}_{1}}{(2\pi)^{3}}P_{\rm m}(k_{1},z)P_{\rm m}(|\mathbf{k}-\mathbf{k}_{1}|,z) \nonumber \\
& \times & \left[ \frac{k(k-2k_{1}\mu)(1-\mu^{2})}{k^{2}_{1}(k^{2}+k^{2}_{1}-2k k_{1}\mu)} \right], \label{eq:P-per}
\end{eqnarray}
is the B-mode power spectrum in Eq.~(\ref{eq:Cell-Tp}). The upper script ``T'' means kSZ temperature fluctuation, and ``p'' means the momentum field. The $f_{\rm gas,i}=f_{\rm gas}(z_{i})$ is $f_{\rm gas}$ function evaluate at redshift $z_{i}$. We believe that $f_{\rm gas}(z)$ is a slow-varying function on average of all scales, so we take the medium value out of the integral in Eq.~(\ref{eq:Cell-Tp}).

Then the two point correlation function at redshift bin $z_{i}$ is
\begin{eqnarray}
\xi^{\rm Tp}(\theta) &=&  \left \langle \Delta_{i} (\mathbf{\hat{n}}) \tilde{\Delta}_{i} (\mathbf{\hat{n}}')  \right \rangle_{\theta=\mathbf{\hat{n}}\cdot \mathbf{\hat{n}'}} \nonumber \\
&=& \sum_{\ell}\left(\frac{2 \ell +1}{4 \pi} \right)C_{\ell}(\chi_{i})P_{\ell}(\cos \theta)B_{\ell}B'_{\ell} \label{eq:xi-Tp} , 
\end{eqnarray}
where $B_{\ell}$ and $B'_{\ell}$ refer to the beam function of the CMB map and reconstructed momentum map. We usually cross-correlate the two maps with the same angular resolution so we normally set $B_{\ell}=B^{\prime}_{\ell}$.

\begin{figure*}[tbp!]
\centerline{
\includegraphics[width=3.4in]{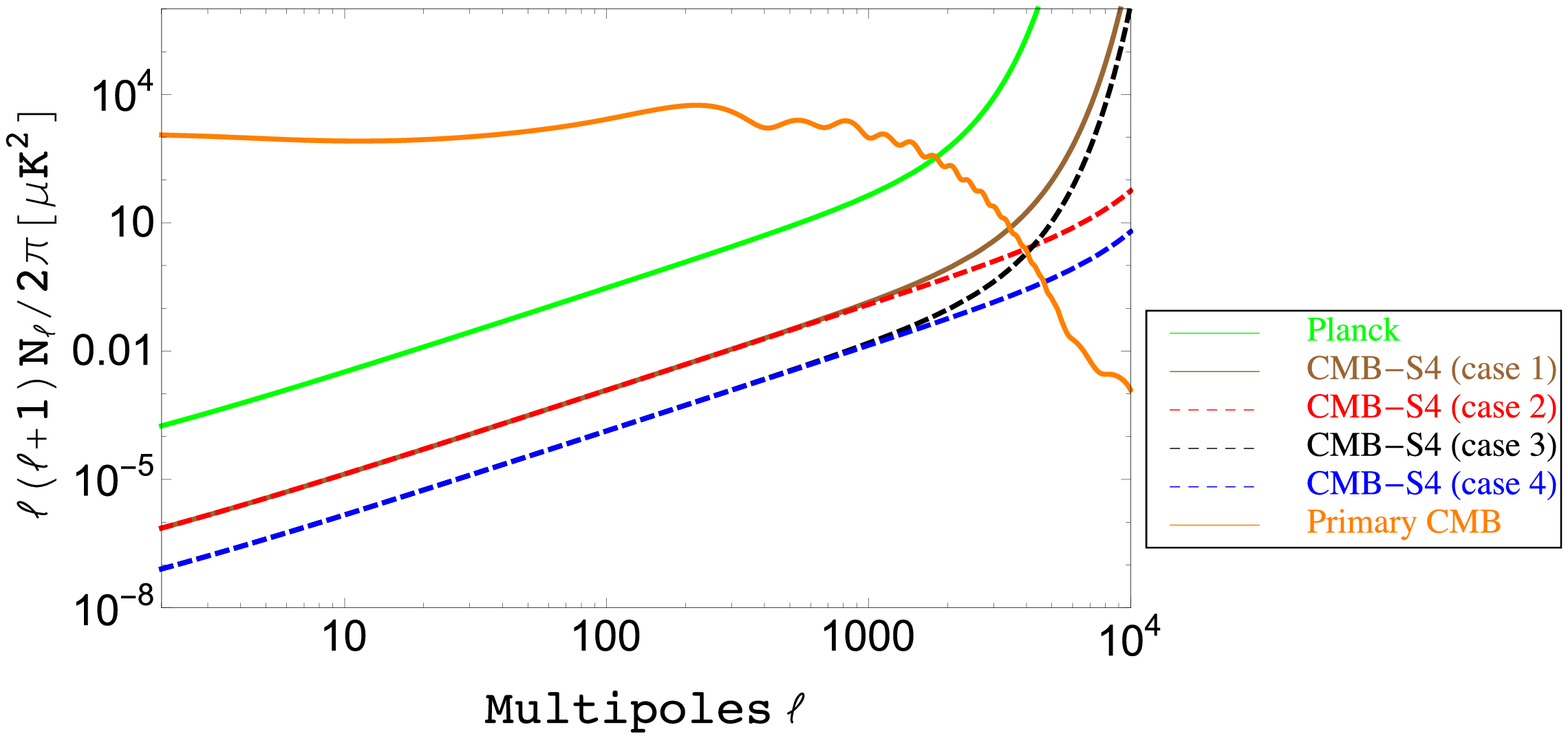}
\includegraphics[width=3.4in]{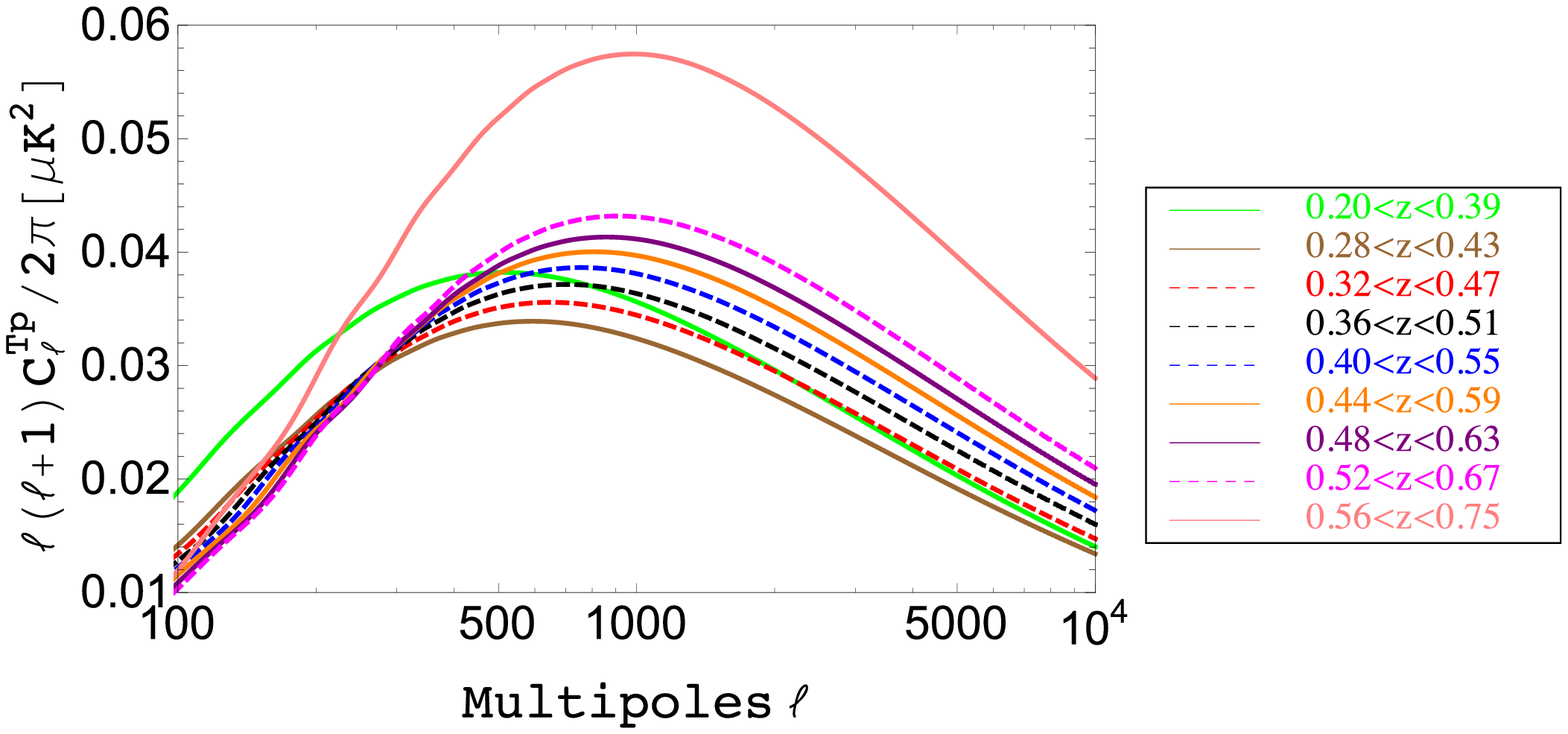}}
\caption{{\it Left}-- the noise power spectrum for {\it Planck} and CMB-S4 experiments listed in Table~\ref{tab-cmb}. We also plot the primary CMB as another source of noise for kSZ. {\it Right}--The kSZ--reconstructed momentum field cross-correlation power spectrum at different redshift bins (with assumption $f_{\rm gas} \equiv 1$). Here we neglect the negative sign in Eq.~(\ref{eq:Cell-Tp}).} \label{fig:cell}
\end{figure*}

\begin{figure}[tbp!]
\centerline{
\includegraphics[width=3.2in]{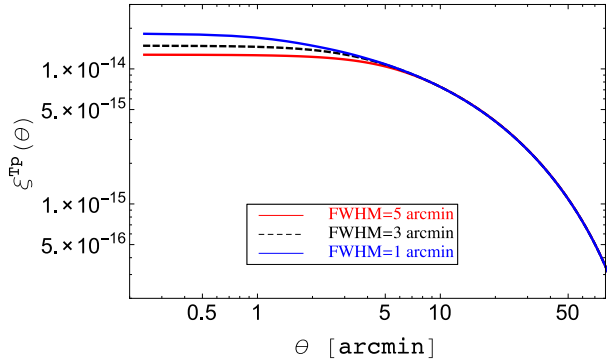}}
\caption{The angular cross-correlation function $\xi^{\rm Tp}(\theta)$ for the redshift bin 1 with $z_{\rm eff}=0.31$.} \label{fig:xi}
\end{figure}

In addition to the cross-correlation power spectrum (Eq.~(\ref{eq:Cell-Tp})), the auto-correlation power spectra of kSZ and momentum field are
\begin{eqnarray}
C^{\rm TT}_{\ell}(\chi_{i}) &=& \frac{1}{2} f^{2}_{\rm gas,i} \left(\frac{\sigma_{\rm T}\chi_{\rm e}\rho_{\rm cr}\Omega_{\rm b}}{\mu_{\rm e}m_{\rm p}c} \right)^{2} \left(\frac{W_{\rm kSZ}(\chi_i)}{\chi_{i}} \right)^{2}  \nonumber \\
& \times &
 \int^{\chi_{i}+\Delta \chi_{i}/2}_{\chi_{i}-\Delta \chi_{i}/2} \der \chi P_{\rm B}\left(\frac{\ell +1/2}{\chi} \right) \label{eq:Cell-TT} \\
C^{\rm pp}_{\ell}(\chi_{i}) &=& \frac{1}{2} \left(\frac{\sigma_{\rm T}\chi_{\rm e}\rho_{\rm cr}\Omega_{\rm b}}{\mu_{\rm e}m_{\rm p}c} \right)^{2} \left(\frac{W_{\rm kSZ}(\chi_i)}{\chi_{i}} \right)^{2}  \nonumber \\
& \times &
 \int^{\chi_{i}+\Delta \chi_{i}/2}_{\chi_{i}-\Delta \chi_{i}/2} \der \chi P_{\rm B}\left(\frac{\ell +1/2}{\chi} \right) \label{eq:Cell-pp}  
\end{eqnarray}
respectively.

\subsection{$f_{\rm gas}$ model}
\label{sec:fgas-model}
We now present the four model of the ionized gas as a function of redshift. These $f_{\rm gas}$ parameter should be understood as the fraction of baryons which are in the status of gas, not yet collapsed into stars and galaxies. We consider the following four models:

\begin{enumerate}

\item Constant $f_{\rm gas}$ through the history of the Universe.

\item Following \citet{Goldberg99} and \citet{Waerbeke14}, $f_{\rm gas} \propto a$, therefore, the second model is
\begin{eqnarray}
f_{\rm gas}(z)=\frac{f_{\rm gas,0}}{(1+z)},
\end{eqnarray}
where $f_{\rm gas,0}$ is the $f_{\rm gas}$ fraction at redshift zero.

\item We add some further variation by allowing the redshift-dependent index to vary, i.e.
\begin{eqnarray}
f_{\rm gas}(z)=\frac{f_{\rm gas,0}}{(1+z)^{\alpha}},
\end{eqnarray}
so we have two parameters here $f_{\rm gas,0}$ and $\alpha$.

\item We allow $f_{\rm gas}$ at each redshift to be different, i.e. allowing the whole $f_{\rm gas}(z)$ function to vary.

\end{enumerate}

\subsection{Fisher matrix}
The cumulative signal-to-noise ratio of the $f_{\rm gas}(z)$ model provides an efficient way of measuring the amount of baryons in form of gas, which provides a measurement of the amount of baryons. At a given redshift $z_{i}$, Fisher matrix for any parameters $\alpha$ and $\beta$ is
\begin{eqnarray}
F^{i}_{\alpha \beta}=\frac{f_{\rm sky}}{2}\sum_{\ell}(2\ell +1)\left(\frac{C^{\rm Tp}_{\ell}(\chi_{i})}{\partial \alpha} \right)\left(M^{-1}_{\ell} \right) \left(\frac{C^{\rm Tp}_{\ell}(\chi_{i})}{\partial \beta} \right), \nonumber \\
\label{eq:Fisher}
\end{eqnarray}
where $f_{\rm sky}$ is the fraction of the sky that are overlapped by CMB and spectroscopic surveys. The covariance matrix
\begin{eqnarray}
M_{\ell}=\hat{C}^{\rm TT}_{\ell}\hat{C}^{\rm pp}_{\ell}+ \hat{C}^{\rm Tp}_{\ell}\hat{C}^{\rm Tp}_{\ell},
\end{eqnarray}
where $C^{\rm TT}_{\ell}$, $C^{\rm pp}_{\ell}$ and $C^{\rm Tp}_{\ell}$ are the kSZ auto-power spectrum, reconstructed momentum field auto-power spectrum, and the kSZ--momentum field cross-power spectrum. Spectra with hat ($\hat{C}_{\ell}$) is the measured power spectrum which essentially contain noise. Since the instrumental noise of CMB does not correlate with the noise in the reconstructed momentum field,
\begin{eqnarray}
\hat{C}^{\rm Tp}_{\ell}=C^{\rm Tp}_{\ell}.
\end{eqnarray}
The kSZ measured power spectrum
\begin{eqnarray}
\hat{C}^{\rm TT}_{\ell}=C^{\rm TT}_{\ell}+N^{\rm TT}_{\ell}+C^{\rm CMB}_{\ell},
\end{eqnarray}
where $C^{\rm CMB}_{\ell}$ is the lensed primary CMB temperature power spectrum which is an essential contamination of the kSZ power spectrum, 
\begin{eqnarray}
N^{\rm TT}_{\ell}= \Delta_{\rm T}^{2}{\rm e}^{\ell^{2}\sigma^{2}_{\rm b}} \label{eq:Nell},
\end{eqnarray}
is the thermal noise of CMB map, $\sigma_{\rm b}=\theta_{\rm FWHM}/\sqrt{8 \ln2}=0.00742\left(\theta_{\rm FWHM}/1^{\circ} \right)$, where $\theta_{\rm FWHM}$ is the beam full-width half maximum (FWHM). The detail values of experimental parameters are listed in Table~\ref{tab-cmb}. Note that since our power spectra (Eqs.~(\ref{eq:Cell-Tp}), (\ref{eq:Cell-TT}) and (\ref{eq:Cell-pp})) are dimensionless, $C^{\rm CMB}_{\ell}$ and $N^{\rm TT}_{\ell}$ need to be normalized with CMB monopole $T_{\rm CMB}=2.725\,$K.

\begin{table}
\begin{centering}
%\arraystretch{1.4}
%\begin{tabular}{|c|c|c|c|}
\begin{tabular}{|c|c|c|}
\hline
CMB experiments & beam FWHM & Effective noise \\ 
& [arcmin] & $\Delta_{\rm T}$ [$\mu$K-arcmin] \\ 
\hline
{\it Planck} & $5$ & $47$  \\ \hline
%%Advanced ACTPol & $1.4$ & $10$ \\ \hline
CMB-S4 (case 1) & $3$ & $3$ \\ \hline
CMB-S4 (case 2) & $1$ & $3$ \\ \hline
CMB-S4 (case 3) & $3$ & $1$ \\ \hline
CMB-S4 (case 4) & $1$ & $1$ \\ \hline
\end{tabular}%
\caption{The specification of CMB experiments used in the forecast. The effective noise means the residual noise in the CMB map after component separation. For the CMB Stage-4 experiments, the specification has not yet determined so we try a few cases for illustrative purposes. These cases were also used in~\cite{Ferraro16,Hill16}.}
\label{tab-cmb}
\end{centering}
\end{table}

For the reconstructed momentum map, the shot noise is much smaller than the reconstructed momentum spectrum, as demonstrated with numerical simulation in~\cite{Shao11}, so we regard
\begin{eqnarray}
\hat{C}^{\rm pp}_{\ell}=C^{\rm pp}_{\ell}.
\end{eqnarray}
%where $C^{\rm pp}_{\ell}$ is the momentum power spectrum, and
%\begin{eqnarray}
%N^{\rm pp}_{\ell}=\frac{1}{N^{\rm 2D}},
%\end{eqnarray}

\begin{table*}
\begin{centering}
%\begin{tabular}{|c|c|c|c|c|c|c|c|}
\begin{tabular}{|c|c|c|c|c|c|c|}
\hline
Bin No. & Redshift Range & Effective $z$ & $N_{\rm NGC}$ & $N_{\rm SGC}$ & $N_{\rm tot}$ & $n_{\rm tot}$  %%& $N^{\rm pp}_{\ell} \times 10^{5}$
\\ \hline
bin 1 & $0.20<z<0.39$ & $0.31$ & $176,899$ & $75,558$ & $252,457$  & $29.9$ \\ \hline %% & $1.02$ \\ \hline
bin 2 & $0.28<z<0.43$ & $0.36$ & $194,754$ & $81,539$ & $276,293$  & $32.7$ \\ \hline %%& $0.93$ \\ \hline
bin 3 & $0.32<z<0.47$ & $0.40$ & $230,388$ & $93,825$ & $324,213$  & $38.4$ \\ \hline %%& $0.79$ \\ \hline
bin 4 & $0.36<z<0.51$ & $0.44$ & $294,749$ & $115,029$ & $409,778$  & $48.5$ \\ \hline %%& $0.63$ \\ \hline
bin 5 & $0.40<z<0.55$ & $0.48$ & $370,429$ & $136,117$ & $506,546$  & $60.0$ \\ \hline %%& $0.51$ \\ \hline
bin 6 & $0.44<z<0.59$ & $0.52$ & $423,716$ & $154,486$ & $578,202$  & $68.5$ \\ \hline %%& $0.44$ \\ \hline
bin 7 & $0.48<z<0.63$ & $0.56$ & $410,324$ & $149,364$ & $559,688$  & $66.3$ \\ \hline %%& $0.46$ \\ \hline
bin 8 & $0.52<z<0.67$ & $0.59$ & $331,067$ & $121,145$ & $452,212$ & $53.6$ \\ \hline %%& $0.57$ \\ \hline
bin 9 & $0.56<z<0.75$ & $0.64$ & $231,505$ & $86,576$ & $318,081$  & $37.7$ \\ \hline %%& $0.81$\\ \hline
\end{tabular}%
\caption{The number and number density distribution at different redshift bins. The columns 4--6 are the number of samples in NGC, SGC and total. The column 7 is the number density of samples per square degree. } %%The column 8 is the equivalent shot noise.}
\label{tab-sample}
\end{centering}
\end{table*}

\section{Observational data}
\label{sec:obs}

\subsection{CMB surveys}

In Table~\ref{tab-cmb}, we show the current {\it Planck} and future CMB Stage-4 experimental specifications. CMB Stage-4 has not yet completely set up so we use a few hypothetical cases as shown in~\cite{Ferraro16,Hill16}. One can see that overall the beam of CMB-S4 experiments will be smaller, and the effective noise in terms of $\mu$K per arcmin will become much smaller than {\it Planck}.

In the left panel of Fig.~\ref{fig:cell}, we plot the noise power spectra for the {\it Planck} and four cases of CMB-S4 experiments, we also plot the lensed primary CMB signal as an additional contaminated component of kSZ. One can see that the lensed CMB has much higher amplitude than most of the current and future CMB experiments from $\ell=2$ to $\ell \simeq 3000$. For the $\ell \gg 3000$ the instrumental thermal noise starts to kick in and become the dominated noise. The beam effect in Eq.~(\ref{eq:Nell}) makes the noise exponentially large at high $\ell$, which restrict the constraining power of high-$\ell$ modes.

\subsection{Spectroscopic survey}
\label{sec:spectro}
Since we aim to cross-correlate the kSZ map from {\it Planck} and future CMB-S4 surveys, we need to cross-correlate it with the reconstructed momentum field from spectroscopic surveys. In~\cite{Planck16-unbound,Carlos}, we have done similar studies. We used the $110,437$ ``Central Galaxies'' from SDSS/DR7 catalogues with stellar mass $\log(M_{\ast}/M_{\odot})>11$ to reconstruct the peculiar velocity field from density field through the continuity equation
\begin{eqnarray}
\frac{\partial \delta}{\partial t} + \nabla \cdot \mathbf{v}=0.
\end{eqnarray}
Then we project the velocity field onto the line-of-sight direction $(\mathbf{v}\cdot \mathbf{\hat{n}})$ to obtain the radial velocities. What we want to do for SDSS-III catalogue and future survey is similar. We need to obtain the Fourier mode density field $\delta(\mathbf{k})$, and then the velocity field $\mathbf{v}(k)$, then reconvert into real space $\mathbf{v}(\mathbf{x})$, and eventually obtain the momentum field.

In Table~\ref{tab-sample}, we list the SDSS-III DR12 samples at different redshift bins. Column 2 is the redshift range, and column 3 is the effective redshift in each bin. The redshift range of this sample is $0.2<z<0.75$, and it contains $\sim 865,000$ and $330,000$ galaxies in the North Galactic Cap (NGC) ($\sim 5900 {\rm deg}^{2}$) and South Galactic Cap (SGC) ($\sim 2500 {\rm deg}^{2}$) respectively. These correspond to the $f_{\rm sky}$ factor to be $f^{\rm NGC}_{\rm sky}=0.14$, $f^{\rm SGC}_{\rm sky}=0.06$, and in total $f_{\rm sky}=0.2$. The final column $n_{\rm tot}$ is the number of samples per unit area within each redshift bin.

\section{Results}
\label{sec:results}

\begin{figure}[tbp!]
\centerline{
\includegraphics[width=3.2in]{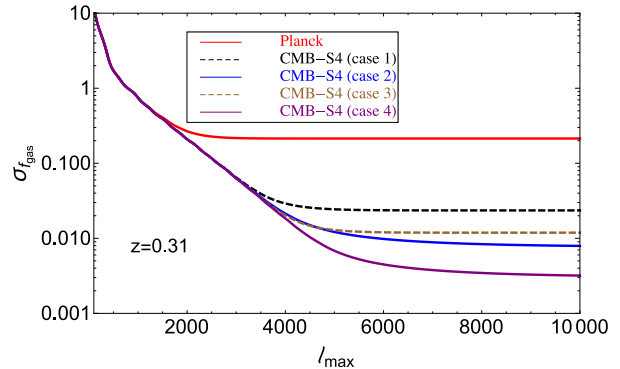}}
\caption{The error of $f_{\rm gas}$ parameter for the first redshift bin ($z=0.31$) as a function of $\ell_{\rm max}$. Red solid, black dashed, blue solid, brown dashed, and purple solid lines are representing different experiments as indicated in the legend.} \label{fig:sigma_fgas}
\end{figure}

\begin{figure*}[tbp!]
\centerline{
\includegraphics[width=3.2in]{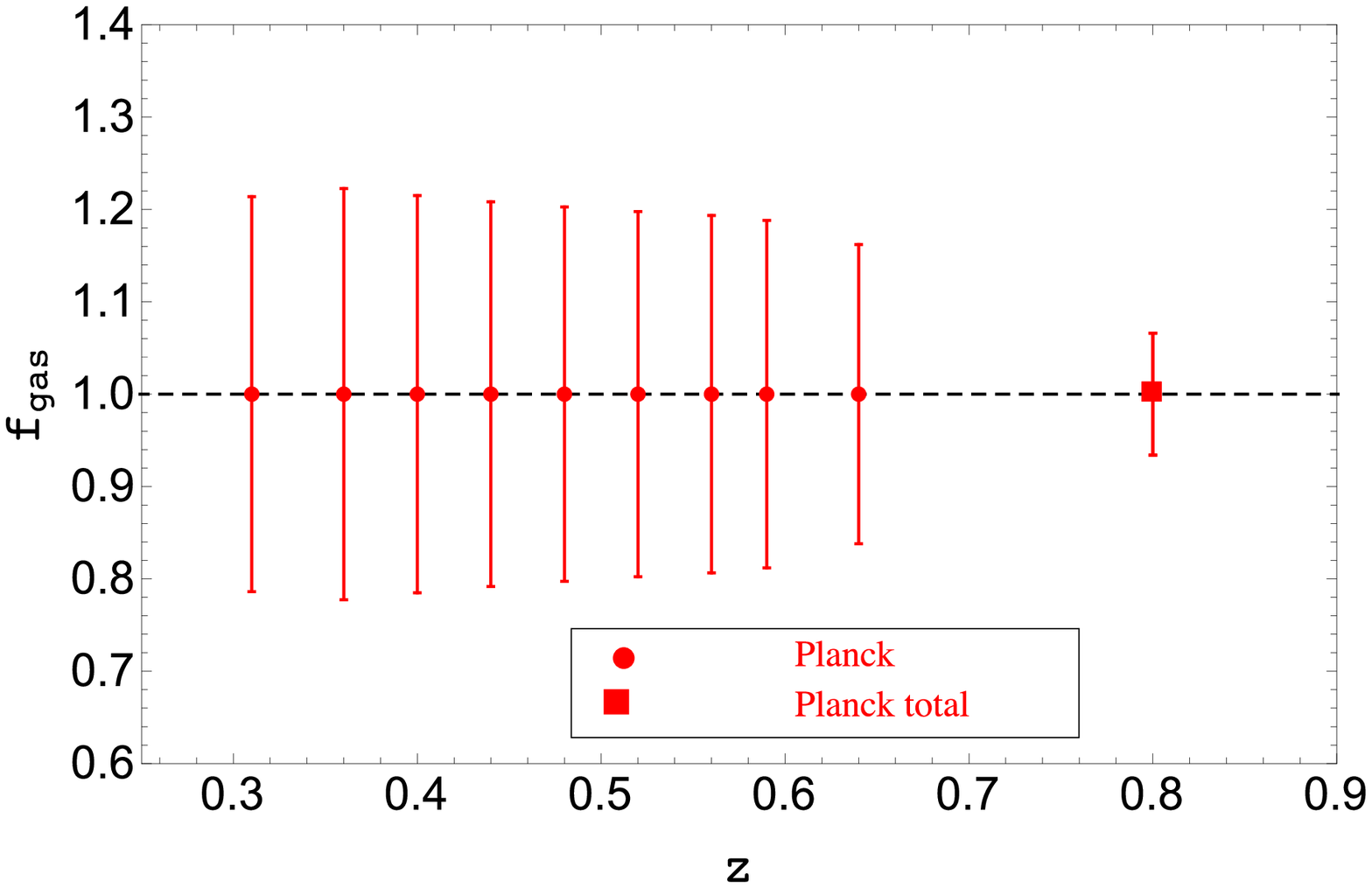}}
\centerline{\includegraphics[width=3.2in]{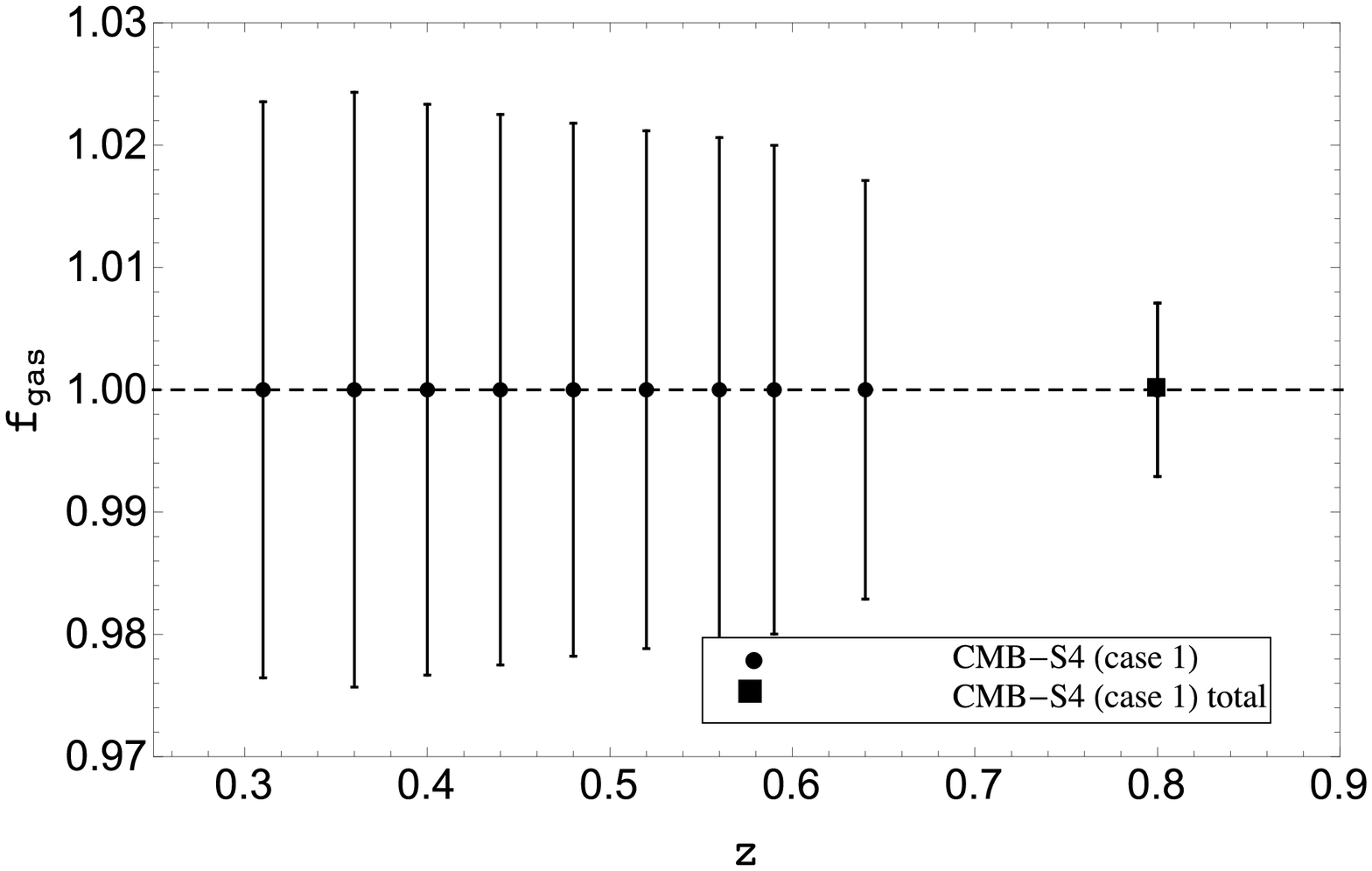}
\includegraphics[width=3.2in]{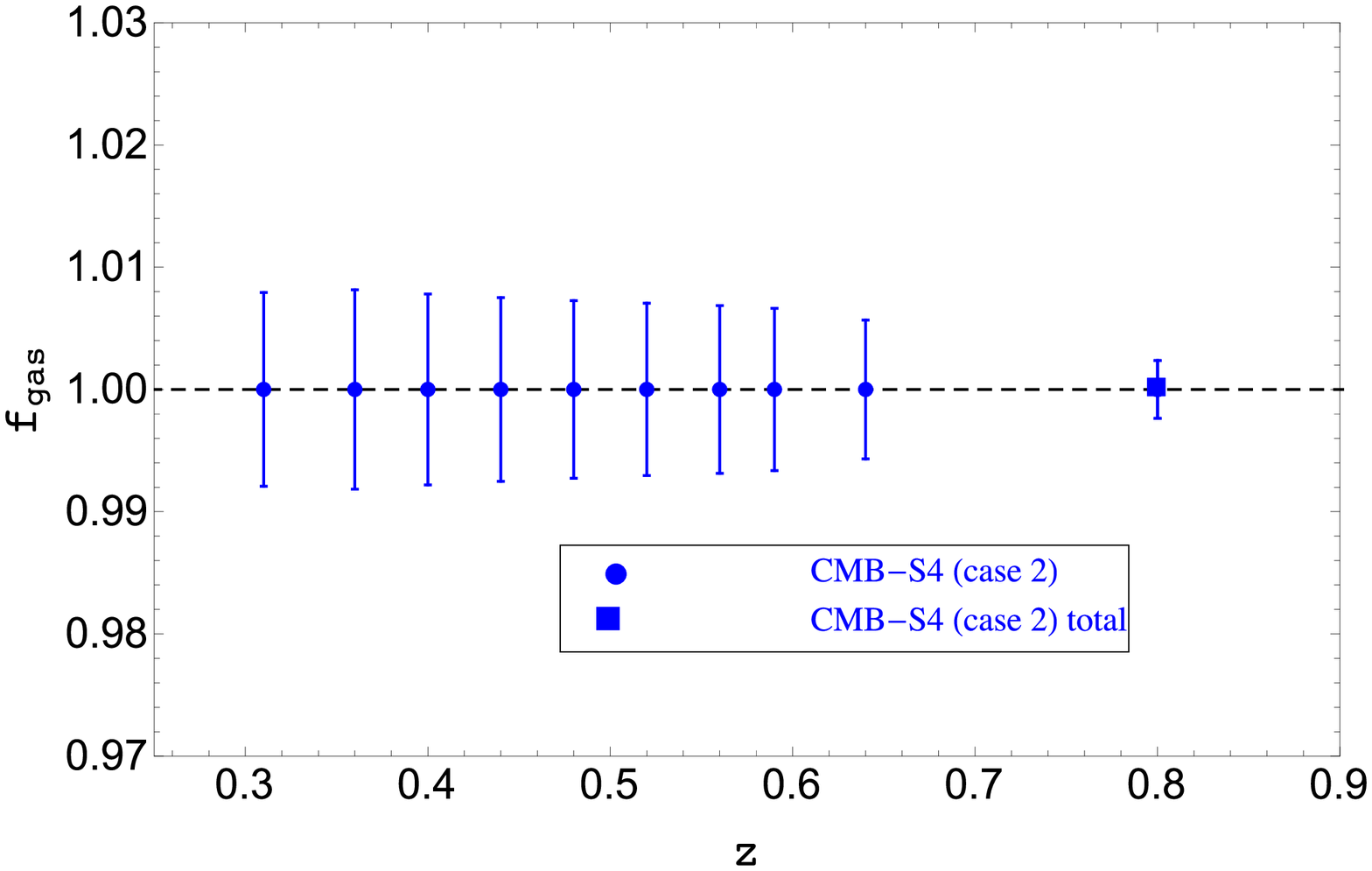}}
\centerline{\includegraphics[width=3.2in]{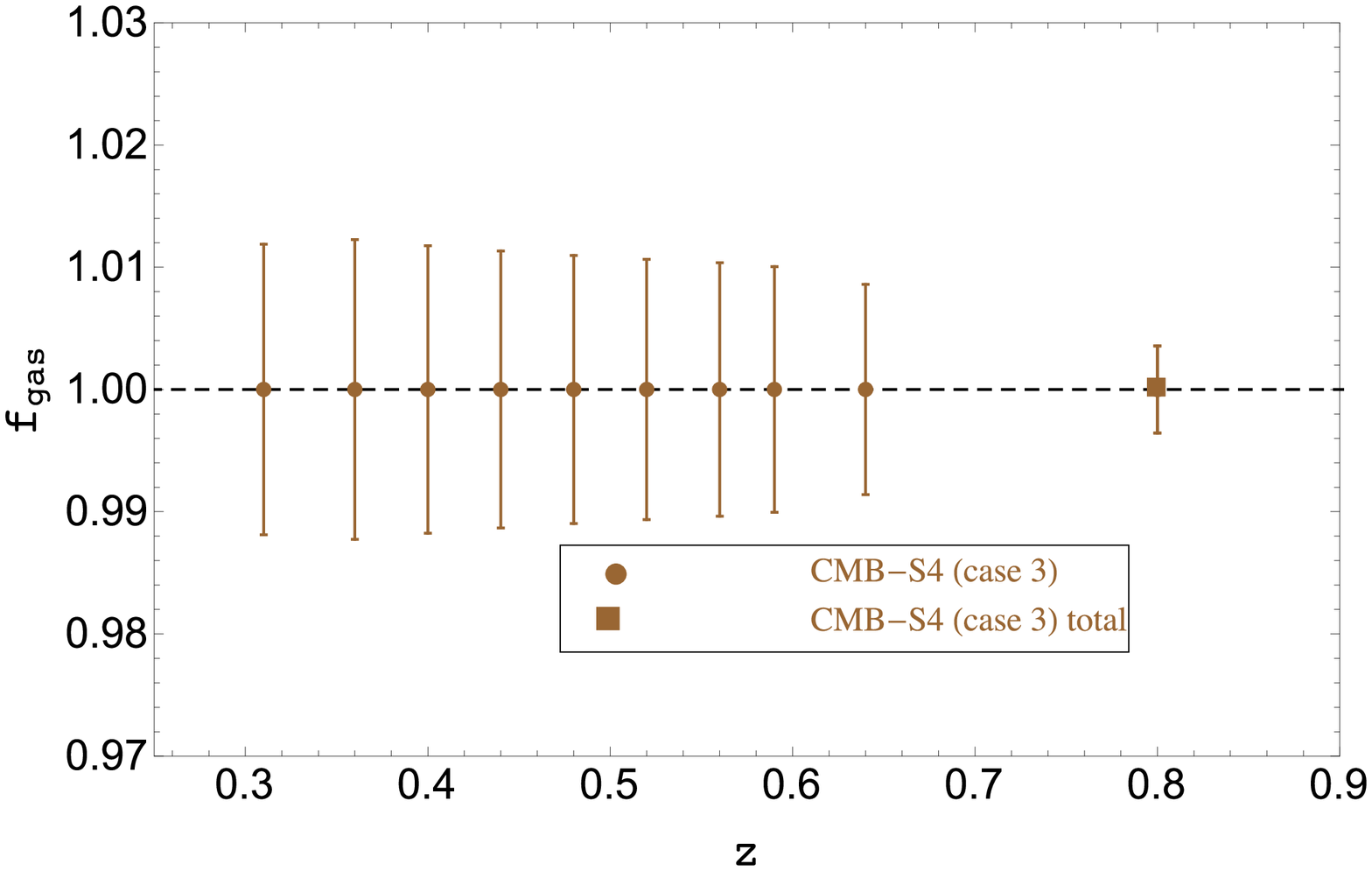}
\includegraphics[width=3.2in]{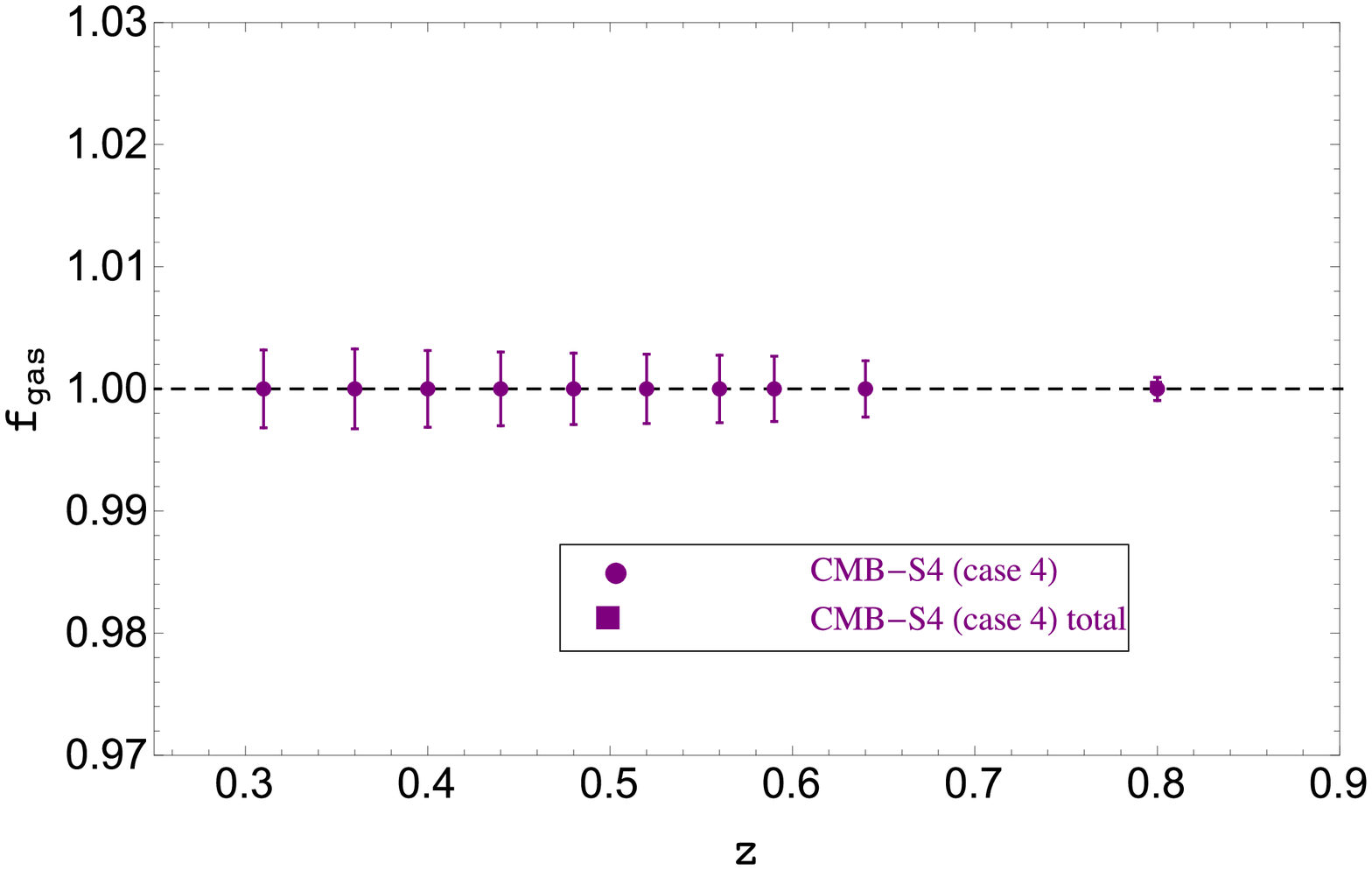}}
\caption{Constraints on $f_{\rm gas}$ (constant model) at different redshift bins from different experiments. The last right data point at $z \simeq 0.8$ is to use all data at each redshift bin to constrain $f_{\rm gas}$, and it is {\it not} the constraint at $z\simeq 0.8$. Note that the scale in the upper panel ({\it Planck}) is much larger than the other four panels.} \label{fig:fgas-const}
\end{figure*}

\begin{figure}[tbp!]
\centerline{
\includegraphics[width=3.2in]{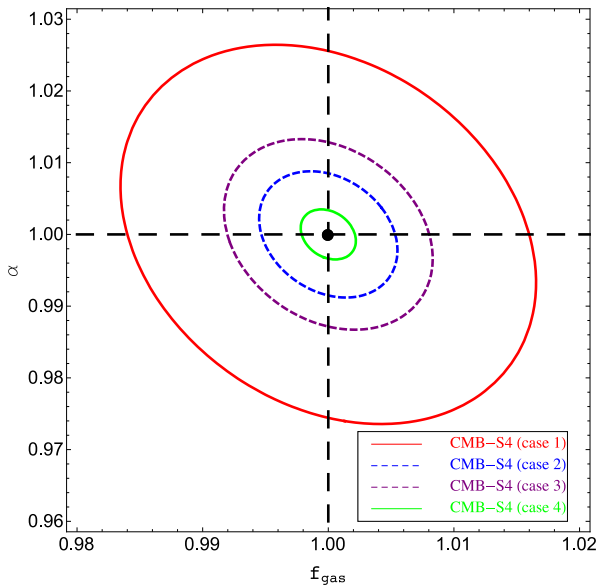}}
\caption{Forecasted joint constraints on the parameters ($f_{\rm gas,0},\alpha$) of the $f_{\rm gas}$ evolution model. The contours indicate the 68\% ($1\sigma$) confidence levels. The input models are indicated by the black point ($f_{\rm gas,0}=1\,,\alpha=1$).} \label{fig:fgas-alp}
\end{figure}

\begin{table*}
\begin{centering}
%\arraystretch{1.4}
%\begin{tabular}{|c|c|c|c|}
\begin{tabular}{|c|c|c|c|c|c|}
\hline
$\sigma_{f_{\rm gas}}$ & {\it Planck} & CMB-S4 (case 1) & CMB-S4 (case 2) & CMB-S4 (case 3) & CMB-S4 (case 4)  \\ \hline
bin 1 &  $0.21$      &  $2.35\times 10^{-2}$ &  $7.92\times 10^{-3}$ &  $1.18\times 10^{-2}$ &  $3.19\times 10^{-3}$ \\ \hline
bin 2 & $0.22$      &  $2.43\times 10^{-2}$ &  $8.15\times 10^{-3}$ &  $1.23\times 10^{-2}$ &  $3.27\times 10^{-3}$ \\ \hline
bin 3 & $0.21$      &  $2.33\times 10^{-2}$ &  $7.81\times 10^{-3}$ &  $1.18\times 10^{-2}$ &  $3.14\times 10^{-3}$ \\ \hline
bin 4 & $0.21$      &  $2.25\times 10^{-2}$ &  $7.52\times 10^{-3}$ &  $1.13\times 10^{-2}$ &  $3.02\times 10^{-3}$ \\ \hline
bin 5 & $0.20$      &  $2.18\times 10^{-2}$ &  $7.27\times 10^{-3}$ &  $1.10\times 10^{-2}$ &  $2.92\times 10^{-3}$ \\ \hline
bin 6 &  $0.20$     &  $2.12\times 10^{-2}$ &  $7.05\times 10^{-3}$ &  $1.06\times 10^{-2}$ &  $2.84\times 10^{-3}$ \\ \hline
bin 7 &  $0.19$     &  $2.06\times 10^{-2}$ &  $6.86\times 10^{-3}$ &  $1.04\times 10^{-2}$ &  $2.76\times 10^{-3}$ \\ \hline
bin 8 &  $0.19$     &  $2.00\times 10^{-2}$ &  $6.64\times 10^{-3}$ &  $1.00\times 10^{-2}$ &  $2.67\times 10^{-3}$ \\ \hline
bin 9 &  $0.16$     &  $1.71\times 10^{-2}$ &  $5.68\times 10^{-3}$ &  $8.61\times 10^{-3}$ &  $2.30\times 10^{-3}$ \\ \hline
Total & $6.60\times 10^{-2}$ &  $7.09\times 10^{-3}$ &  $2.36\times 10^{-3}$ &  $3.57\times 10^{-3}$ &  $9.52\times 10^{-4}$ \\ \hline 
\end{tabular}%
\caption{The error of constant $f_{\rm gas}$ model at each redshift bin for five different CMB experiments. The last row is the resulted error if we treat $f_{\rm gas}$ is a constant throughout evolution history.}
\label{tab:error-fgas-const}
\end{centering}
\end{table*}

\begin{table*}
\begin{centering}
%\arraystretch{1.4}
%\begin{tabular}{|c|c|c|c|}
\begin{tabular}{|c|c|c|c|c|c|}
\hline
$\sigma_{f_{\rm gas,0}}$ & {\it Planck} & CMB-S4 (case 1) & CMB-S4 (case 2) & CMB-S4 (case 3) & CMB-S4 (case 4)  \\ \hline
bin 1 &  $0.28$      &  $3.08\times 10^{-2}$ &  $1.04\times 10^{-2}$ &  $1.56\times 10^{-2}$ &  $4.15\times 10^{-3}$ \\ \hline
bin 2 & $0.30$      &  $3.31\times 10^{-2}$ &  $1.11\times 10^{-2}$ &  $1.66\times 10^{-2}$ &  $4.42\times 10^{-3}$ \\ \hline
bin 3 & $0.30$      &  $3.27\times 10^{-2}$ &  $1.09\times 10^{-2}$ &  $1.64\times 10^{-2}$ &  $4.35\times 10^{-3}$ \\ \hline
bin 4 & $0.30$      &  $3.24\times 10^{-2}$ &  $1.08\times 10^{-2}$ &  $1.63\times 10^{-2}$ &  $4.31\times 10^{-3}$ \\ \hline
bin 5 & $0.30$      &  $3.22\times 10^{-2}$ &  $1.07\times 10^{-2}$ &  $1.62\times 10^{-2}$ &  $4.28\times 10^{-3}$ \\ \hline
bin 6 &  $0.30$     &  $3.21\times 10^{-2}$ &  $1.07\times 10^{-2}$ &  $1.61\times 10^{-2}$ &  $4.26\times 10^{-3}$ \\ \hline
bin 7 &  $0.30$     &  $3.21\times 10^{-2}$ &  $1.07\times 10^{-2}$ &  $1.61\times 10^{-2}$ &  $4.25\times 10^{-3}$ \\ \hline
bin 8 &  $0.30$     &  $3.17\times 10^{-2}$ &  $1.05\times 10^{-2}$ &  $1.59\times 10^{-2}$ &  $4.19\times 10^{-3}$ \\ \hline
bin 9 &  $0.27$     &  $2.80\times 10^{-2}$ &  $9.28\times 10^{-3}$ &  $1.40\times 10^{-2}$ &  $3.70\times 10^{-3}$ \\ \hline
Total & $9.79\times 10^{-2}$ &  $1.05\times 10^{-2}$ &  $3.51\times 10^{-3}$ &  $5.29\times 10^{-3}$ &  $1.40\times 10^{-3}$ \\ \hline 
\end{tabular}%
\caption{The error of $f_{\rm gas,0}$ of the model $f_{\rm gas}(z)=f_{\rm gas,0}/(1+z)$ model at each redshift bin for five different CMB experiments. The last row is the resulted error if we combine all data at different redshift bins.}
\label{tab:error-fgas-1plusz}
\end{centering}
\end{table*}

\begin{table}
\begin{centering}
%\arraystretch{1.4}
%\begin{tabular}{|c|c|c|c|}
\begin{tabular}{|c|c|c|}
\hline
Observation & $\sigma_{f_{\rm gas,0}}$ & $\alpha$ \\ \hline
{\it Planck} & $0.53$        & $1.34$ \\ \hline    
CMB-S4 (case 1) &  $ 5.77\times 10^{-2}$ & $0.14$ \\ \hline     
CMB-S4 (case 2) &  $1.92 \times 10^{-2}$ & $4.83\times 10^{-2}$ \\ \hline
CMB-S4 (case 3) &  $2.90\times 10^{-2}$  & $7.28\times 10^{-2}$  \\ \hline
CMB-S4 (case 4) &  $7.69 \times 10^{-3}$ & $1.93\times 10^{-2}$ \\ \hline
\end{tabular}%
\caption{The forecasted errors of $f_{\rm gas,0}$ and $\alpha$ for the model $f_{\rm gas}(z)=f_{\rm gas,0}/(1+z)^{\alpha}$ evolution. The fiducial $f_{\rm gas,0}=1$ and $\alpha=1$.}
\label{tab:error-fgas-1plusz-alp}
\end{centering}
\end{table}

\subsection{Power spectrum and correlation function}
\label{sec:curve}

In the right panel of Fig.~\ref{fig:cell}, we plot the kSZ--momentum field angular power spectrum for the $9$ redshift bins. This power spectrum is calculated via Eq.~(\ref{eq:Cell-Tp}) by assuming $f_{\rm gas}=1$. One can see that the cross-correlation power spectra at different redshift bins have slightly different amplitudes and shapes, but overall the amplitudes peak at $\ell \simeq 1000$ and with amplitude roughly $\ell(\ell+1)C^{\rm Tp}_{\ell}/2\pi \simeq 0.05\,{\mu}{\rm K}^{2}$. Comparing to the left panel of Fig.~\ref{fig:cell}, at $\ell \sim 1000$ regime, the primary CMB signal starts to drop and the instrumental noise of {\it Planck} starts to rise. However, for the four cases of CMB-S4 experiments the instrumental noises are still quite low comparing to the primary CMB on scales of $\ell \simeq 1000$, and they only start to rise up over primary CMB at $\ell \gtrsim 3000$.

In this paper we did not discuss the component separation method, but in reality, one needs to separate out the lensed primary CMB and thermal SZ effect to obtain the kSZ map. Separating the thermal SZ effect needs to apply a frequency space filter, which is a mature technique developed in~\cite{Remazeilles11}.
%the series of {\it Planck} papers~\cite{Planck-comp1,Planck-comp2}. 
The {\tt NILC}, {\tt SEVEM}, {\tt SMICA}, and {\tt COMMANDER} are the foreground cleaned maps with different component separation techniques~\cite{Planck-comp1,Planck-comp2}. These maps suppress Galactic foreground and dust but keep the CMB and kSZ signal, since the kSZ has very little spectral distortion to the primary CMB. To further separate the kSZ from the CMB, one needs to apply the spatial filter to the map, such as ``aperture photometry method'' developed in~\cite{Planck14-vel,Planck16-unbound} or the matched filter technique~\cite{Tegmark98,Ma13}. So the final map will not only depend on the instrumental noise of the CMB map but also the residual foreground separation. For this reason, we keep the lensed primary CMB as a source of noise which contaminates the observed kSZ map. In reality, this term can be successfully suppressed if we apply a suitable filter to the map.

In Fig.~\ref{fig:xi}, we plot the angular correlation function $\xi^{\rm Tp}(\theta)$ for the redshift bin 1, $0.20<z<0.39$, for different beam sizes. One can see that the difference between convolution with different beam sizes only exists at the very central angular scales of the cross-correlated map. This is because since the cross-correlation function peaks at $\ell \sim 1000$ (right panel in Fig.~\ref{fig:cell}), this corresponds to the angular scales of $\sim 20\,$arcmin. This is larger than the CMB FWHM beam function. So once the kSZ angular correlation function convolves with the CMB beam through Eq.~(\ref{eq:xi-Tp}) the only difference will be at small angular scales, while the general amplitude and shape do not change very much. 

Once the real data is available, one can work either in real space by calculating the correlated data points and covariance matrix, or work in the angular space by transforming the data into $\ell$-space, and then fit the angular power spectrum. The results should be equivalent to each other. In below, we work in the $\ell$-space for the forecast.

\subsection{Results of the forecast}
\label{sec:forecast}

Here we present the results of the forecasts on four models of $f_{\rm gas}(z)$ listed in Sect.~\ref{sec:fgas-model}. 

\begin{enumerate}

\item $f_{\rm gas}=$constant model (free parameter: $f_{\rm gas}$). In this model the $f_{\rm gas}$ is a constant throughout cosmic time. We use the Fisher matrix (Eq.~(\ref{eq:Fisher})) to calculate the $F_{f_{\rm gas}f_{\rm gas}}$ for each redshift bin and then inverse it to obtain the error $\sigma_{f_{\rm gas}}=F_{f_{\rm gas}f_{\rm gas}}^{-1/2}$. In Fig.~\ref{fig:sigma_fgas}, we plot the cumulative error by adding $\ell$-modes from $\ell=2$ to $\ell_{\rm max}$ for the first redshift bin $z_{\rm eff}=0.31$. One can see that the cumulative error starts to decrease as more high-$\ell$ modes are added in. But the error will be saturated and does not decrease any more once it reaches some critical $\ell$ mode. For {\it Planck} survey the saturated scale is $\ell \sim 2000$. This is because at $\ell \sim 2000$ (Fig.~\ref{fig:cell} left panel), the instrumental noise starts to blow up and becomes dominated noise sources. Therefore, for $\ell \gtrsim 2000$ modes they are not able to pin down the error of $f_{\rm gas}$ parameter. However, such situation is changed if one uses CMB-S4 experiment, because their noise level and FWHM is smaller than {\it Planck}. From Fig.~\ref{fig:sigma_fgas} one can see that $\sigma_{f_{\rm gas}}$ can continuously go down as more and more higher $\ell$-modes are added into the constraints. For the very optimistic CMB-S4 case 4 experiment with $\Delta_{\rm T}=1\,{\mu}\,{\rm K}$-arcmin and $\theta_{\rm FWHM}=1\,$arcmin, the error of $f_{\rm gas}$ can be as low as $\sim 3.9 \times 10^{-3}$ for the first redshift bin constraint.

In Fig.~\ref{fig:fgas-const}, we show the constraints on the $f_{\rm gas}$ parameter at each redshift bin, and the joint constraint by adding the constraints from each redshift bin together. For the joint constraint, in order to compare it with constraints from each redshift bin, we plot it at $z \sim 0.8$. Note that the error in the upper panel for {\it Planck} measurement is much larger than the other four panels for CMB-S4 experiments. We list the detail values of error in Table~\ref{tab:error-fgas-const}. One can see that for each redshift bin, the errors from {\it Planck}, CMB-S4 case 1, 2, 3, 4 are $\sim 0.2$, $2\times 10^{-2}$, $7 \times 10^{-3}$, $1 \times 10^{-2}$ and $2 \times 10^{-3}$ respectively. The errors of constraints from all redshift bins are $6.6 \times 10^{-2}$ for {\it Planck}, and $\sim 10^{-3}$ for CMB-S4 experiments cases 1--3, and $\sim 10^{-4}$ for CMB-S4. This will become a very powerful result to constrain the gaseous evolution.

\item $f_{\rm gas}(z)=f_{\rm gas,0}/(1+z)$ model (free parameter: $f_{\rm gas,0}$).

Similarly, we calculate the Fisher matrix by considering the redshift evolution $(1+z)$ factor. The forecasted error of $f_{\rm gas,0}$ is similar to the case of $f_{\rm gas}=$constant model, but with slightly larger error-bars. We listed the results of $\sigma_{\rm gas,0}$ in Table~\ref{tab:error-fgas-1plusz}. By comparing Table~\ref{tab:error-fgas-1plusz} with Table~\ref{tab:error-fgas-const}, one can see that the errors of $f_{\rm gas,0}$ combined from different redshift bins are slightly broadened to be $\sim 0.1$, $1.05 \times 10^{-2}$, $3.51 \times 10^{-2}$, $5.29 \times 10^{-2}$, $1.40 \times 10^{-2}$ for {\it Planck} and CMB-S4 cases 1--4 respectively.

\item $f_{\rm gas}(z)=f_{\rm gas,0}/(1+z)^{\alpha}$ model (free parameters: $f_{\rm gas,0}$ and $\alpha$). We further release the index of redshift evolution $\alpha$ as a free parameter in the forecast. We calculate the $2\times 2$ Fisher matrix where the off-diagonal term is $F_{f_{\rm gas,0}\, \alpha}$. Then the marginalized error of any parameter ($\theta$) is just the $\sigma_{\theta}=\left(F^{-1}\right)_{\theta \theta}$.

In Table~\ref{tab:error-fgas-1plusz-alp}, we presented the marginalized errors of $f_{\rm gas,0}$ and $\alpha$ parameters for the five experimental cases by assuming fiducial ($f_{\rm gas,0}=1$ and $\alpha=1$). Each value is the joint constraints from all 9 redshift bins. By comparing Table~\ref{tab:error-fgas-1plusz-alp} with the last row of Table~\ref{tab:error-fgas-1plusz}, one can see that the errors become bigger due to the inclusion of additional $\alpha$ parameters. In addition, the constraint on $\alpha$ from {\it Planck} is too weak to distinguish the redshift evolution of $f_{\rm gas}$ model, but for CMB-S4 experiments, they will be able to constrain the redshift evolution very well.

We further calculate the joint constraints of the $f_{\rm gas,0}$ and $\alpha$ parameters by calculating the effective $\chi^{2}$
\begin{eqnarray}
\chi^{2}=\left(f_{\rm gas,0}-f^{\rm fid}_{\rm gas,0} \right)C^{-1} \left(\alpha-\alpha^{\rm fid} \right),
\end{eqnarray}
where $C=F^{-1}$, and $f^{\rm fid}_{\rm gas,0}=1$, $\alpha^{\rm fid}=1$. We neglect the case of {\it Planck} since it is unable to provide a reasonable constraint. But for CMB-S4 four cases shown in Fig.~\ref{fig:fgas-alp}, one can see that the joint constraints can become tighter and tighter once the thermal noise and beam size becomes smaller. For case 4 of CMB-S4, it will provide stringent test on the $f_{\rm gas}(z)$ model.

\item Variable $f_{\rm gas}(z)$ model. 

Finally, we examine how well one can test the variable $f_{\rm gas}(z)$ model. This means that at each redshift $f_{\rm gas}$ value is different, but it is a slow-varying function. In Table~\ref{tab:error-fgas-const}, we forecast the error of $f_{\rm gas}(z)$ model at each different redshift bin. One can see that if $f_{\rm gas}(z)$ is slow-varying, {\it Planck} may only be able to constrain its value at error of $\sim 0.2$ level, whereas for four cases of CMB-S4 experiments, the errors can be pined down to $2 \times 10^{-2}$, $7 \times 10^{-3}$, $1 \times 10^{-2}$, $2 \times 10^{-3}$ respectively. The small error achievable by CMB-S4 experiment makes it is possible to constrain and rule out rapid variable of $f_{\rm gas}(z)$ models.

\end{enumerate}

\section{Conclusion}
\label{sec:conclude}

In this paper, we have examined prospective measurement of the ionized gas model with the current and future CMB experiments cross-correlated with spectroscopic survey. We calculate the cross-correlation between the kSZ map obtainable from CMB survey ({\it Planck} and CMB-S4 survey) with reconstructed momentum maps from SDSS-III survey. We first construct the template of reconstructed momentum map from spectroscopic survey, and then calculate the cross-correlation power spectrum between the two. We then use Fisher matrix technique to forecast the error of the parameters of interests in $f_{\rm gas}(z)$ model.

We find that for the constant $f_{\rm gas}$ model, {\it Planck} survey is able to constrain it to $6.6 \times 10^{-2}$ level ($1\sigma$ CL), while the CMB-S4 experiments are able to constrain it till $\mathcal{O}(10^{-3})$ level ($1\sigma$ CL). For the very optimistic CMB-S4 case 4 where the $\theta_{\rm FWHM}=1\,$arcmin and instrumental noise $\Delta_{\rm T}=1\,{\mu}{\rm K}$-arcmin, then constraint can reach as $\sigma_{f_{\rm gas}} \lesssim 10^{-3}$ ($1\sigma$ CL).

We then examine the two other cases of redshift evolution, $f_{\rm gas}(z)=f_{\rm gas,0}/(1+z)$ and $f_{\rm gas}(z)=f_{\rm gas,0}/(1+z)^{\alpha}$. One can see that for $f_{\rm gas}(z)=f_{\rm gas,0}/(1+z)$ model the constraints on $f_{\rm gas,0}$ parameter is slightly broadened but still very tight. The signal-to-noise can reach $10$ for {\it Planck} and $100$--$700$ for four cases of CMB-S4 experiments. For $f_{\rm gas}(z)=f_{\rm gas,0}/(1+z)^{\alpha}$ model, {\it Planck} itself cannot constrain the index $\alpha$ very well, but for the various cases of CMB-S4 experiment, the detection of $f_{\rm gas,0}$ can reach $17$--$130$ and the error of $\alpha$ can be pined down to $\mathcal{O}(0.1)$ and even $0.02$. This will provide a stringent constraint on the evolution of ionized gas in galaxy formation.

Finally, we discuss the prospective constraints on $f_{\rm gas}(z)$ evolution model, i.e. a slow-varying function of $f_{\rm gas}$ as a function of redshift. We find that {\it Planck} can only constrain its value down to level of $\sigma_{f_{\rm gas}} \simeq 0.2$, but for various CMB-S4 experimental cases, the error can be lowered down till $\mathcal{O}(10^{-3})$.

In conclusion, in terms of constraining the parameters of gas evolution, tomographic kSZ method with cross-correlation of momentum field is clearly a very powerful tool and will likely overtake the other methods in searching for the missing baryons. In this paper, we have not directly considered the issue of model selection of the $f_{\rm gas}(z)$ models. However, it is likely that, in addition to constraining model parameters, future sensitive CMB-S4 observations will also allow us to distinguish between models of ionized gas evolution such as those considered in this paper.

\textit{Acknowledgements}-- 
We are grateful for discussion with Carlos Hern\'{a}ndez-Monteagudo, Mathieu Remazeilles, Yuting Wang, Xiao-dong Xu, Pengjie Zhang, Gong-Bo Zhao. This work is 
supported by National Research Foundation of South Africa with 
grant no.105925.

\appendix

\section{Power spectrum $C^{\rm Tp}_{\ell}$}
\label{sec:power-cell}
\subsection{Limber cancellation}

In this section, we calculate the power spectrum of kSZ--momentum field cross-correlation $C^{\rm Tp}_{\ell}$. From Eqs.~(\ref{eq:deltaTT2}) and (\ref{eq:delta-i-n}), the projected momentum effect is a line-of-sight integral of the momentum vector. For any vector, it can be decomposed into a curl-free (gradient part) and a curl- (divergence-free) part. Therefore, the momentum function $\mathbf{p}$ can be decomposed into a curl-free or a gradient part $\mathbf{p}_{\rm E}$ satisfying $\nabla \times \mathbf{p}_{\rm E}=0$ and a divergence-free or curl part $\mathbf{p}_{\rm B}$ satisfying $\nabla \cdot \mathbf{p}_{\rm B}=0$, i.e. $\mathbf{p}=\mathbf{p}_{\rm E}+\mathbf{p}_{\rm B}$. Now it is easy to show that the gradient part $\mathbf{p}_{\rm E}$ does not contribute to the kSZ effect, as long as the comoving distance of the integral $\Delta \chi \gg 100\,h^{-1}$Mpc. We know that the integral of $\mathbf{p}_{\rm E}$'s contribution is
\begin{eqnarray}
\frac{\Delta T}{T} &\sim & \int_{\Delta \chi} \der \chi W_{\rm kSZ}(\chi)(\mathbf{\hat{n}}\cdot \mathbf{p}_{\rm E}) \nonumber \\
&=& \int_{\Delta \chi} \der \chi W_{\rm kSZ}(\chi)(\mathbf{\hat{n}}\cdot \vec{\nabla}\phi(\chi)) \nonumber \\
& \sim & \int \der \chi W_{\rm kSZ}(\chi) \mathbf{\hat{n}}\cdot \vec{\nabla}\left[\der^{3} \mathbf{k} \tilde{\phi}(\mathbf{k}){\rm e}^{-i\mathbf{k}\cdot \mathbf{r}} \right] \nonumber \\
& = & \int \der^{3}\mathbf{k} (-i) (\mathbf{k}\cdot \mathbf{\hat{n}}) \left[ \int_{\Delta \chi} \der \chi W_{\rm kSZ}(\chi) \tilde{\phi}(\mathbf{k}) {\rm e}^{i\mathbf{k}\cdot \mathbf{r}} \right], \nonumber \\ 
\end{eqnarray}
where $W_{\rm kSZ}(z)=(1+z)^{2}{\rm e}^{-\tau(z)}$. Since at low redshift, $W_{\rm kSZ}(\chi)$ is a slow-varying function of the $\chi$, then as long as your integration range $\Delta \chi$ is much greater than the coherent length of the velocity field potential ($\sim 100\,h^{-1}$Mpc), the square bracket integration always rapidly oscillates (exponential function is indeed a cosine function) and only the small value of $\mathbf{k}\cdot \mathbf{\hat{n}}$ contributes to the integral. But because of the pre-factor, such integral is highly suppressed. Pictorially, for the gradient term, the $\mathbf{k}$ is parallel to $\mathbf{\hat{n}}$, so while integrating over a long range, the contributions from troughs and crests of each Fourier model cancel approximately when projected along the line-of-sight. However, there are two exceptions that are incomplete cancellation: (1) super-horizon ``dark flows'', since the coherent length of such dark flow is Gpc, there are fewer troughs and crests; (2) Normal galaxy flow but at the epoch of reionization, where the patchy reionization causes $W_{\rm kSZ}(\chi)$ to vary significantly over $k\simeq 10\,h\,{\rm Mpc}^{−1}$ scales~\cite{Shao11}.

\subsection{Angular power spectrum}
Therefore, all we need to calculate is $\mathbf{\hat{n}}\cdot \mathbf{p}_{\rm B}$. From Eq.~(\ref{eq:delta-i-n}), we know that (we neglect the constant prefactor at the moment, see also~\cite{Kaiser92})
\begin{eqnarray}
\Delta a_{\ell m} &=& W_{\rm kSZ}(\chi_{0}) (4\pi i^{\ell}) \int^{\chi_{i}+\Delta \chi_{i} /2}_{\chi_{i}- \Delta \chi_{i} /2} \der \chi \nonumber \\
& \times &
\left(\int  \frac{\der^{3}\mathbf{k}}{(2\pi)^{3}} (\mathbf{\hat{n}}\cdot \mathbf{p}_{\rm B}(\mathbf{k})) j_{\ell}(k\chi)Y_{\ell m}(\mathbf{\hat{k}})  \right). 
\end{eqnarray}
Then we have
\begin{eqnarray}
\langle \Delta a_{\ell m} \Delta a_{\ell' m'} \rangle &=& 
W_{\rm kSZ}^{2}(\chi_{i}) (4\pi)^{2} (i^{\ell}(-i)^{\ell'}) \int^{\chi_{i}+\Delta \chi_{i} /2}_{\chi_{i}- \Delta \chi_{i} /2} \der \chi \der \chi' 
\nonumber \\
& \times & \int  \frac{\der^{3}\mathbf{k}}{(2\pi)^{3}} \frac{\der^{3}\mathbf{k}'}{(2\pi)^{3}} \left \langle (\mathbf{\hat{n}}\cdot \mathbf{p}_{\rm B}(\mathbf{k}))
(\mathbf{\hat{n}'}\cdot \mathbf{p}_{\rm B}(\mathbf{k}'))
\right \rangle \nonumber \\
& \times  & j_{\ell}(k\chi) j_{\ell'}(k' \chi') Y_{\ell m}(\mathbf{\hat{k}})Y_{\ell' m'}(\mathbf{\hat{k}'}). \label{eq:delta-alm2}    
\end{eqnarray}

Because
\begin{eqnarray}
&& \left \langle (\mathbf{\hat{n}}\cdot \mathbf{p}_{\rm B}(\mathbf{k}))
(\mathbf{\hat{n}'}\cdot \mathbf{p}_{\rm B}(\mathbf{k}'))
\right \rangle \nonumber \\
&& = \sum_{ij} \mathbf{\hat{n}}_{i}\mathbf{\hat{n}}'_{j}\left \langle q_{\rm B, i}(\mathbf{k}) q_{\rm B, j}(\mathbf{k}')
\right \rangle \nonumber \\
&& = \sum_{ij} \mathbf{\hat{n}}_{i}\mathbf{\hat{n}}'_{j}\frac{1}{2}\left(\delta_{ij}-\mathbf{\hat{k}}_{i} \mathbf{\hat{k}}_{j} \right) \left \langle \mathbf{p}_{\rm B}(\mathbf{k}) \cdot \mathbf{p}_{\rm B}(\mathbf{k}') \right \rangle \nonumber \\
&& = \frac{1}{2}\left(\mathbf{\hat{n}} \cdot \mathbf{\hat{n}}' -(\mathbf{\hat{n}}\cdot \mathbf{\hat{k}}) (\mathbf{\hat{n}}'\cdot \mathbf{\hat{k}}) \right) \left \langle \mathbf{p}_{\rm B}(\mathbf{k}) \cdot \mathbf{p}_{\rm B}(\mathbf{k}') \right \rangle \nonumber \\
& & = \frac{1}{2}\left(\mathbf{\hat{n}} \cdot \mathbf{\hat{n}}' -(\mathbf{\hat{n}}\cdot \mathbf{\hat{k}}) (\mathbf{\hat{n}}'\cdot \mathbf{\hat{k}}) \right) (2\pi)^{3}\delta^{3}_{\rm D}(\mathbf{k}-\mathbf{k}')P_{\rm B}(k).\nonumber \\
\end{eqnarray}
Then we substitute this equation back to Eq.~(\ref{eq:delta-alm2}). Note that here we use two assumption: 
\begin{itemize}

\item Small angle approximation, so that the angle between the two line-of-sight is small, therefore $\mathbf{\hat{n}}\cdot \mathbf{\hat{n}}'=1$.

\item The wavenumber $\mathbf{k}$ is perpendicular to the line-of-sight, so $\mathbf{\hat{n}}\cdot \mathbf{k}=0$. 

\end{itemize}
Then we have

\begin{eqnarray}
\left \langle \Delta a_{\ell m} \Delta a^{\ast}_{\ell' m'} \right\rangle =C_{\ell} (\chi_{i}) \delta_{\ell \ell'} \delta_{m m'},
\end{eqnarray}
where (we put back the constant pre-factor)
\begin{eqnarray}
C_{\ell}(\chi_{i}) &=& -\frac{1}{2} f_{\rm gas,i} \left(\frac{\sigma_{\rm T}\chi_{\rm e}\rho_{\rm cr}\Omega_{\rm b}}{\mu_{\rm e}m_{\rm p}c} \right)^{2} \left(\frac{W_{\rm kSZ}(\chi_i)}{\chi_{i}} \right)^{2}  \nonumber \\
& \times &
 \int^{\chi_{0}+\Delta \chi_{i}/2}_{\chi_{i}-\Delta \chi_{i}/2} \der \chi P_{\rm B}\left(\frac{\ell +1/2}{\chi} \right) . \nonumber \\
\end{eqnarray}

\subsection{Momentum power spectrum}
\label{sec:power}

The rest of the task is to compute $P_{\rm B}(k)$. Remember the $\mathbf{\tilde{p}}(\mathbf{k})$ is 
\begin{eqnarray}
\tilde{\mathbf{p}}(\mathbf{k}) &=& \int \frac{\der^{3}\mathbf{k}_{1}}{(2\pi)^{3}}\tilde{\mathbf{v}}(\mathbf{k}_{1})\delta(\mathbf{k}-\mathbf{k}_{1}) \nonumber \\
&=& i(\dot{a}f) \int \frac{\der^{3}\mathbf{k}_{1}}{(2\pi)^{3}} \left(\frac{\mathbf{\hat{k}}_{1}}{k_{1}} \right) \delta(\mathbf{k}_{1})\delta(\mathbf{k}-\mathbf{k}_{1}), \label{eq:q-k}
\end{eqnarray}
where we have used $f=\der \ln D/\der \ln a$, $\delta(\mathbf{k},z)=a\delta_{0}(\mathbf{k})$. 

We now want to project the Eqs.~(\ref{eq:q-k}) onto the direction perpendicular to $\mathbf{k}$, therefore, we multiply $(\delta_{ij}-k_{i}k_{j}/k^{2})$ onto each component of $\tilde{\mathbf{q}}$, we obtain
\begin{eqnarray}
\tilde{\mathbf{p}}_{\rm B}(\mathbf{k}) = i(\dot{a}f) \int \frac{\der^{3}\mathbf{k}_{1}}{(2\pi)^{3}} \delta(\mathbf{k}_{1})\delta(\mathbf{k}-\mathbf{k}_{1}) \left(\frac{\mathbf{k}_{1}}{k^{2}_{1}} -\frac{(\mathbf{k}\cdot \mathbf{k}_{1})\mathbf{k}}{k^{2}_{1}k^{2}} \right), \nonumber \\
\label{eq:pb-k} 
\end{eqnarray}
therefore from $\langle \tilde{\bf p}_{\rm B}(\mathbf{k}) \tilde{\bf p}_{\rm B}(\mathbf{k}') \rangle =(2\pi)^{3}\delta^{3}(\mathbf{k}-\mathbf{k}')P_{\rm B}(k)$, we have
\begin{eqnarray}
P_{\rm B}\left(k,z \right) &=& (\dot{a}f)^{2} \int \frac{\der^{3}\mathbf{k}_{1}}{(2\pi)^{3}}P_{\rm m}(k_{1},z)P_{\rm m}(|\mathbf{k}-\mathbf{k}_{1}|,z) \nonumber \\
& \times & \left[ \frac{k(k-2k_{1}\mu)(1-\mu^{2})}{k^{2}_{1}(k^{2}+k^{2}_{1}-2k k_{1}\mu)} \right]. \label{eq:P-per1}
\end{eqnarray}

\bibliography{reference}

\end{document}